\documentclass[journal=jpclcd,manuscript=article]{achemso}

\usepackage[version=3]{mhchem} 

\newcommand{\degree}{\ensuremath{^\circ}}

\author{William R. French}
\affiliation[Vanderbilt University]{Department of Chemical and Biomolecular Engineering, Vanderbilt University, Nashville, TN}
\author{Christopher R. Iacovella}
\affiliation[Vanderbilt University]{Department of Chemical and Biomolecular Engineering, Vanderbilt University, Nashville, TN}
\author{Ivan Rungger}
\affiliation[Trinity College Dublin]{School of Physics and CRANN, Trinity College, Dublin 2, Ireland}
\author{Amaury Melo Souza}
\affiliation[Trinity College Dublin]{School of Physics and CRANN, Trinity College, Dublin 2, Ireland}
\author{Stefano Sanvito}
\affiliation[Trinity College Dublin]{School of Physics and CRANN, Trinity College, Dublin 2, Ireland}
\author{Peter T. Cummings}
\email{peter.cummings@vanderbilt.edu}
\affiliation[Vanderbilt University]{Department of Chemical and Biomolecular Engineering, Vanderbilt University, Nashville, TN}
\alsoaffiliation[Oak Ridge National Laboratory]{Center for Nanophase Materials Sciences, Oak Ridge National Laboratory, Oak Ridge, TN}

\title{Structural Origins of Conductance Fluctuations in Gold-Thiolate Molecular Transport Junctions}

\begin{document}

\begin{abstract}

We report detailed atomistic simulations combined with high-fidelity conductance calculations to probe the structural origins of conductance fluctuations in thermally evolving Au-benzene-1,4-dithiolate-Au junctions. We compare the behavior of structurally ideal junctions (electrodes with flat surfaces) to structurally realistic, experimentally representative junctions resulting from break junction simulations. The enhanced mobility of metal atoms in structurally realistic junctions results in significant changes to the magnitude and origin of the conductance fluctuations. Fluctuations are larger by a factor of 2-3 in realistic junctions compared to ideal junctions. Moreover, in junctions with highly deformed electrodes, the conductance fluctuations arise primarily from changes in the Au geometry, in contrast to results for junctions with non-deformed electrodes, where the conductance fluctuations are dominated by changes in the molecule geometry. These results provide important guidance to experimentalists developing strategies to control molecular conductance for device applications, and also to theoreticians invoking simplified structural models of junctions to predict their behavior.  

\end{abstract}

\begin{figure}[h!]  
	\centering
	\includegraphics[width=2.5in]{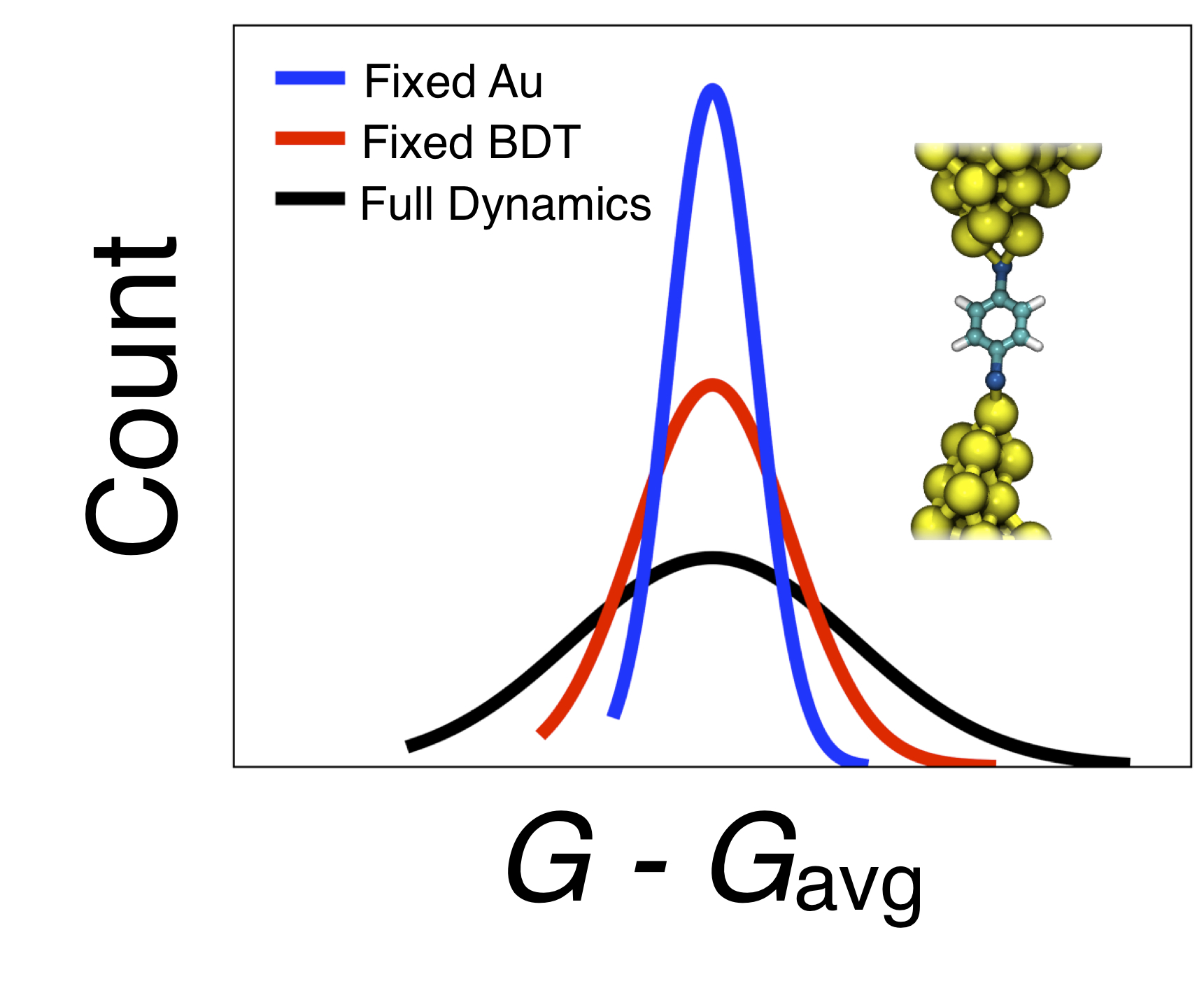}
	\label{fig:toc}
\end{figure}



\pagebreak


Conductance fluctuations in molecular transport junctions are a major barrier to the construction of reliable molecular-based circuitry \cite{McCreery:2009,Ulrich:2006}. The fluctuations arise from changes in the junction structure between successive junction rupture and reformation events or due to thermal motion \cite{Malen:2009}. Therefore, developing strategies to suppress conductance fluctuations relies critically on understanding the structural origins of the fluctuations. For instance, following a study \cite{Venkataraman-Nature:2006} that showed that the conductance through biphenyl molecular wires depends on the dihedral angle between the phenyl rings, Kiguchi et al. \cite{Kiguchi:2012} synthesized a rotaxane structure to limit changes in the dihedral angle, thereby suppressing the conductance fluctuations. For simpler molecules, such as benzene-1,4-dithiolate (BDT), the conductance fluctuations are typically attributed to changes in the metal-molecule contact geometry (bonding site and tilt angle) \cite{Kim:2011,Haiss:2008,Tsutsui:2006}. However, the electrode geometry may play an increasingly important role for systems involving mechanical elongation and deformation of the junction \cite{Ulrich:2006,Malen:2009,Venkataraman-Nature:2006,Kiguchi:2012,Kim:2011,Tsutsui:2006,Tsutsui:2009,Frei:2012,Arroyo:2011,Yokota:2010,Huang:2007,Bruot:2012,Xiao:2004}. For example, Au-thiolate bonding results in significant deformation of the electrodes in break junction experiments. Several groups \cite{Frei:2012,Arroyo:2011,Yokota:2010} have recently investigated the role of Au-thiolate bonding in break junction environments, but the exact structural origins of the conductance behavior remains unclear. Many computational studies have been performed to elucidate these features \cite{Andrews:2008,Kim:2010,Sergueev:2010,Pontes:2011}, however, almost all have considered idealized electode geometries which may not be fully representative of the experimental counterparts. Here, we compare Au-BDT-Au junctions with structurually ideal vs. realistic features to understand (1) the origins of conductance fluctuations and (2) how ideality may influence the conclusions drawn from computation.

To evaluate the effect of junction structure on conductance, we perform atomistic simulations combined with conductance calculations; in this approach, snapshots are periodically extracted from the simulations and then used as input in electron transport calculations. Previous studies \cite{Andrews:2008,Kim:2010} of this kind focused on ideal geometries with a single molecule sandwiched between two flat surfaces. Here, we focus on geometries that are more representative of those likely to appear in widely used break junction experiments \cite{Ulrich:2006,Malen:2009,Venkataraman-Nature:2006,Kiguchi:2012,Kim:2011,Tsutsui:2006,Tsutsui:2009,Frei:2012,Arroyo:2011,Huang:2007,Bruot:2012,Xiao:2004}. In addition to simulations where all atoms are dynamic, we perform two separate simulations (from the same starting point) for each junction where either the BDT geometry or Au geometry is fixed. The BDT geometry (i.e., intramolecular geometry and Au-BDT contact geometry) is fixed by treating the molecule (including the Au atoms covalently linked to the BDT) as a rigid body. In separate simulations, the positions of the Au atoms are fixed while the BDT is free to move. By eliminating specific degrees of freedom within our simulations, we are able to determine the independent contributions of changes in the Au and BDT geometries to the conductance fluctuations.

\begin{figure}[h!]  
	\centering
	\includegraphics[width=6.5in]{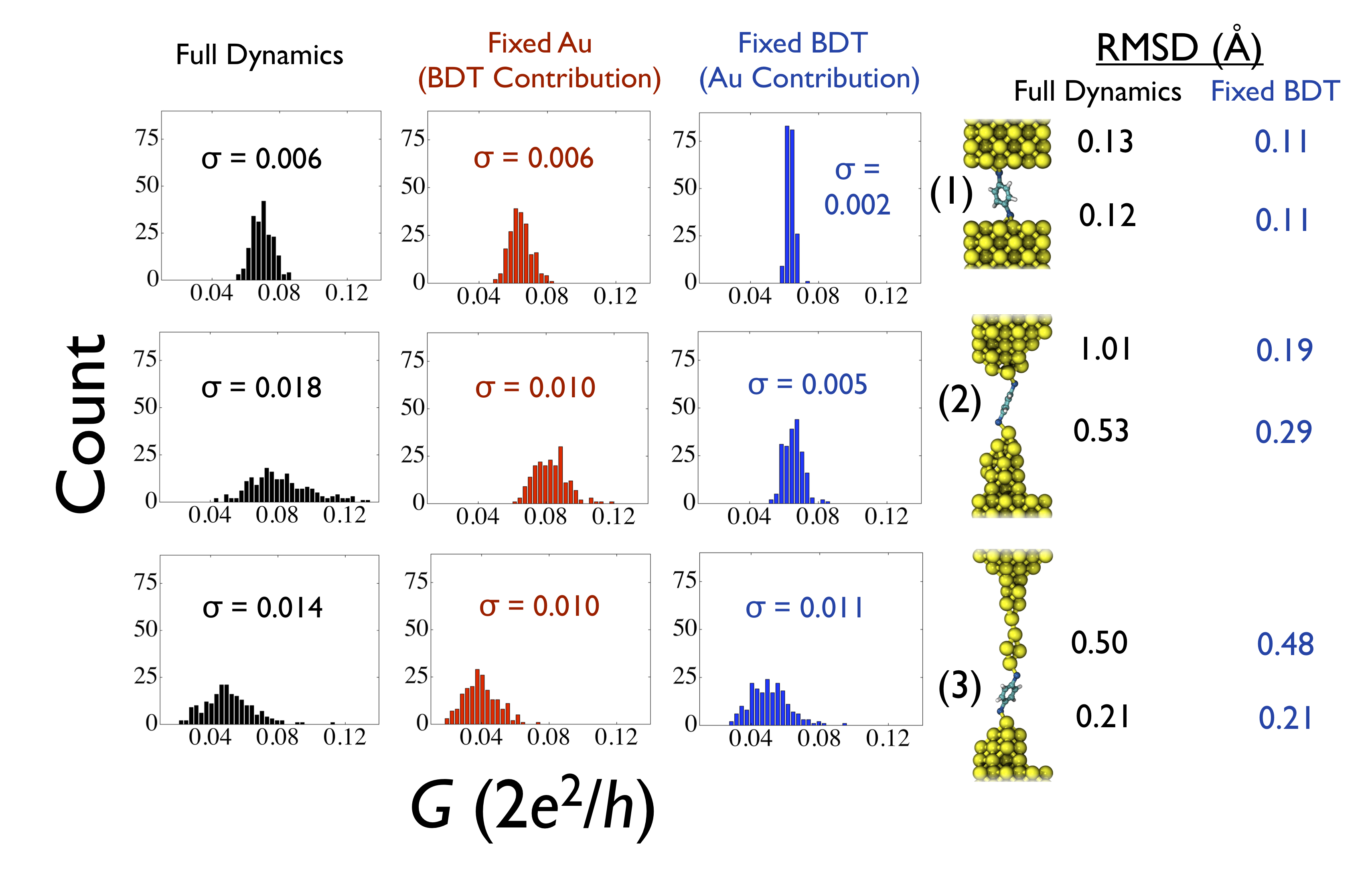}
        \caption{Calculated conductance histograms. (Top row) Ideal, flat-surface junction, (middle row) a junction with curved tips, and (bottom row) a highly deformed junction. For each junction, three separate simulations are run: (left column of plots) one where all atoms in the junction are dynamic, (middle column of plots) one with the Au atomic positions fixed, and (right column of plots) one with fixed BDT geometry. The standard deviation, $\sigma$, is shown with each histogram, and the RMSD of the Au atom bonded with BDT in each tip is shown on the far right.}   
	\label{fig:histograms}
\end{figure}

Figure 1 compares the conductance histograms for the three junctions and three simulation types. We first consider the fully dynamic simulation results, which reveal important differences between the ideal and non-ideal junctions. The ideal junction (junction 1) produces conductance histograms that are much narrower than those for the non-ideal junctions (junctions 2 and 3). The peak width (standard deviation, $\sigma$) is more than an order of magnitude lower than the average conductance for the ideal junction, while for the non-ideal junctions the peak width is on the same order of magnitude as the conductance values themselves. The large peak widths in the non-ideal junctions result from increased geometric freedom, and may present challenges for applications where a device is required to maintain a target conductance value within some threshold. The shape of the conductance histogram also changes for the non-ideal junctions. While the distribution for the ideal junction appears Gaussian (as expected for nonresonant tunneling through molecules\cite{Reuter:2012}), results for the non-ideal junctions exhibit long tails spanning conductance values much higher than the peak values. This results from a transition in the electron transport mechanism from far off resonance in ideal junctions to off resonance in non-ideal junctions (see Supporting Information for a comparison of the transmission curves).

It is apparent from the relative peak widths in the fixed BDT results that the role of the electrodes becomes increasingly important as the electrodes are deformed. For the ideal junction, the Au atoms are closely bound to their lattice sites and thus do not contribute significantly to the conductance fluctuations. In fact, the fully dynamic peak width is completely resolved in the fixed Au simulation, indicating that the conductance fluctuations are dominated by the ability of BDT to explore configuration space. The situation changes for junction 2, as the fully dynamic peak width is not completely resolved from the fixed Au simulation. This suggests that the motion of the electrodes facilitates the sampling of a greater range of contact geometries. For junction 3, the peak widths are similar for the three types of simulations, albeit slightly wider for the fully dynamic simulation, which demonstrates the importance of the interplay between Au and BDT geometry in these systems; that is, changes in the BDT geometry are often enabled by changes in the electrode geometry, and vice versa. Importantly, the peak width in the fixed BDT simulation is larger than that for the fixed Au simulation, indicating a transition in the primary origin of conductance fluctuations from changes in the molecule geometry to changes in the Au geometry. We attribute this transition to the enhanced \textit{dynamic structural fluxionality} \cite{Rashkeev:2007} (lengthening and weakening of Au-Au bonds) in the top tip of junction 3. To support this explanation, we calculate the root-mean-square deviation (RMSD) of the position of the Au atom bonded to BDT relative to its average position. In cases where multiple Au atoms are bonded to BDT, the Au atom that is on average closest to the bonded S atom is considered. As shown in Figure 1, the peak widths scale with the RMSD magnitude. With the BDT geometry fixed, the top tip in junction 3 exhibits the highest RMSD due to its low coordination. In the fully dynamic simulations, junction 2 produces the highest RMSD, which is consistent with its large spread in conductance. The high RMSD results from electrode rearrangements, contributing to the mobility beyond simple fluctuations about a single position.

\begin{figure}[h!]  
	\centering
	\includegraphics[width=4.0in]{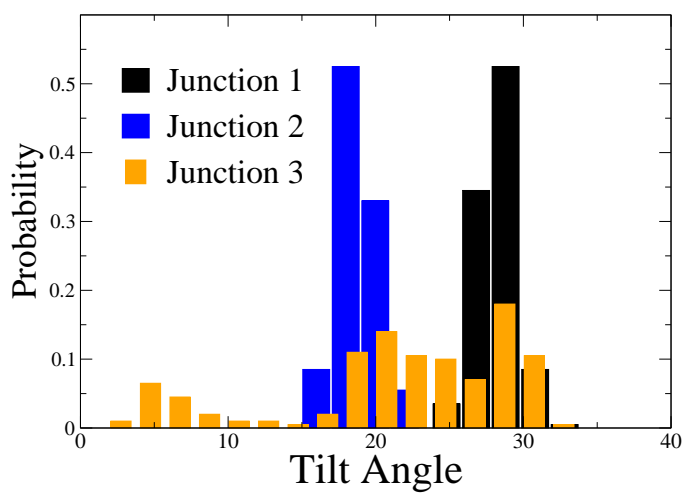}
        \caption{Tilt angle ($\degree$) distribution during the fixed Au simulations.}   
	\label{fig:tilt}
\end{figure}

The sampled molecular tilt angles also change between the different junction geometries. However, increases in the range of sampled tilt angles do not increase conductance fluctuations significantly. Figure 2 plots histograms of the tilt angle (angle between the S-S vector and $z$-axis) during the fixed Au simulations. Note that the distribution is much wider in junction 3 where the molecule can more easily rotate around the ``sharp'' upper tip. It has been shown that in junctions where the electrodes are represented as flat surfaces, the conductance is sensitive to the tilt angle at values greater than 20$\degree$ \cite{Haiss:2008}. It is therefore surprising that the increased tilting freedom of the BDT molecule in junction 3 results in conductance fluctuations that are slightly smaller than those resulting from the fixed BDT simulation.  In the case of sharp tips, the strong relationship between tilt angle and conductance may not apply since the interactions between the molecule and electrode(s) are limited by the small number of metal atoms in the vicinity of the metal-molecule bond(s). In different environments ($e.g.$, higher tilt angles and relatively flat tips) where the carbon atoms in BDT can interact with the Au tips, the range in sampled tilt angles may make more significant contributions to the conductance fluctuations.  

\begin{figure}[h!]  
	\centering
	\includegraphics[width=4.0in]{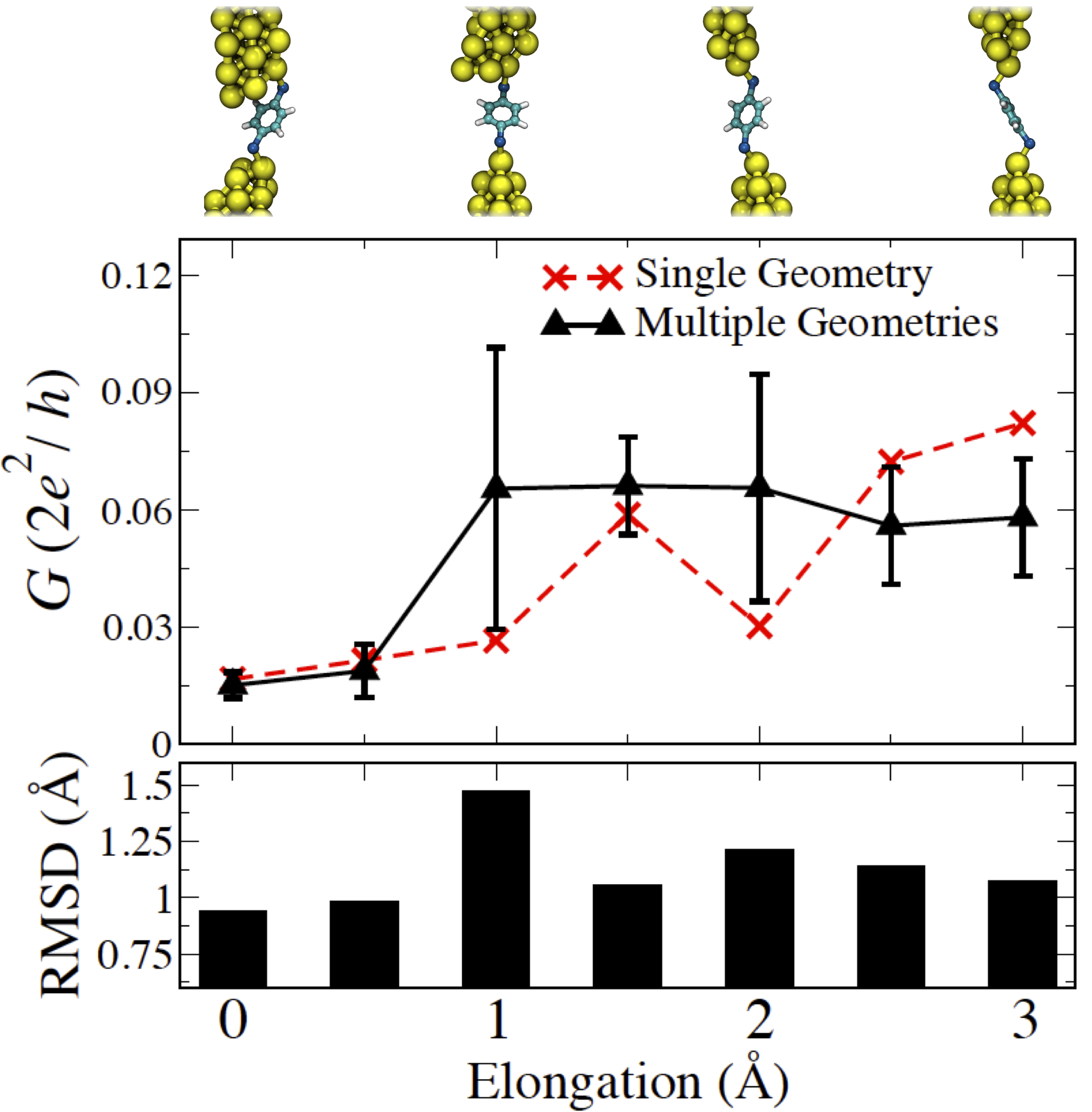}
        \caption{(Top) Thermally averaged and single-geometry conductance trace for Au-BDT-Au junction undergoing elongation. The initial junction geometries are shown above for every \AA\ of elongation. (Bottom) Plot showing the average RMSD of the Au atom bonded to BDT. }   
	\label{fig:elongation}
\end{figure}

An important consideration in molecular break junction experiments is how the conductance fluctuations change as a junction is elongated. Thus, we next explore conductance fluctuations in a junction undergoing mechanical elongation. The junction elongation procedure is described in the Computational Methods section. The average conductance and standard deviation at each elongation length are presented in Figure 3, along with the RMSD value of the BDT-bonded Au atom in each tip during the MD-MC simulation; the conductance from the initial geometry at each elongation length is also plotted for comparison. The error bar sizes in Figure 3 strongly indicate that mechanical deformation of the junction increases the conductance fluctuations, as the error bars at the early stages 0.0 and 0.5 \AA\ are very small, before increasing significantly at 1 \AA. This behavior is dictated by high-mobility structures that form in response to mechanical elongation; that is, the tips become less ideal as the junction is deformed. The fluctuations and RMSD are largest at elongation lengths of 1.0 and 2.0 \AA, where significant structural rearrangements of the electrodes occur during the simulation. Figure 3 also highlights the importance of considering more than a single geometry when calculating conductance, especially for systems where the relative changes in conductance are small between the various structures. It is clear that the single-geometry data fails to capture the average behavior over the entire range. Additionally, Figure 3 illustrates the difficulty in identifying junction structure based on experimental conductance histograms, as different junction structures may have similar average conductance values and highly overlapping distributions.

In summary, we have investigated conductance fluctuations in structurally distinct Au-BDT-Au junctions. We demonstrated that conductance fluctuations in non-ideal junctions are higher than those in an ideal junction. We also showed that while changes in the molecule geometry dominate conductance fluctuations in structurally ideal junctions, the enhanced motion of the Au atoms in deformed electrodes leads to an increased contribution to the conductance fluctuations from changes in the electrode geometry. The minimal role of Au geometry in flat-surface junctions and its significant role in highly deformed junctions highlight the importance of controlling structure in single-molecule conductance measurements. These results also show that conductance fluctuations in thiolate-based break junctions, where significant deformation to the electrodes occurs \cite{Frei:2012,Arroyo:2011,Huang:2007}, may be difficult to control as both the molecule and electrode motion make significant contributions to the fluctuations. The presence of other complex bonding arrangements (e.g., Au-S-Au-S-Au ``staple'' motifs \cite{Cossaro:2008}) at the Au-S interface may further complicate this issue \cite{Hakkinen:2012,Strange:2010}. In contrast, linkers with weaker coupling (e.g., amines) are unlikely to result in the formation of structures such as Au-Au$_{2}$-Au, and thus efforts to control molecular motion \cite{Kiguchi:2012} may prove highly effective for controlling conductance fluctuations. This work also demonstrates the importance of the choice of electrode used in computational studies.
 
\section{Computational Methods}

\hspace*{\parindent} \textit{Hybrid Molecular Dynamics-Monte Carlo Simulations.} In this study, we simulate three structurally distinct junctions: BDT connected between two perfectly flat Au(100) surfaces (junction 1), two curved tips (junction 2), and a curved tip and a Au-Au$_{2}$-Au structure (junction 3). Junctions 2 and 3 were constructed via a hybrid molecular dynamics-Monte Carlo (MD/MC) simulation approach \cite{Pu:2010,French:2012}, and were selected as they represent two extremes of the types of junctions that would be expected in real break junction experiments. Note, the stability and appearance of the Au-Au$_{2}$-Au structure and other similar low-coordination structures are supported by density functional theory calculations \cite{Li:2007,Tavazza:2010}. Each junction is evolved at 77 K (a common temperature in experiment \cite{Tsutsui:2009} and simulation \cite{Andrews:2008}) by performing 200 cycles of MD-MC sampling, where a cycle consists of 0.2 ns of MD followed by 200,000 MC moves; at the end of each cycle, the conductance is computed. The hybrid MD-MC protocol yields ergodic sampling of the possible configurations in each junction; that is, the conductance histograms are converged, implying that we have fully sampled configuration space. In the junction elongation simulations (see Figure 3), elongation is carried out by displacing the upper lead by 0.1 {\AA} in the [001] direction followed by 20 ps of MD and 100,000 MC moves. Every 0.5 {\AA} of elongation a snapshot is extracted and evolved without stretching for 20 MD-MC cycles, with the conductance computed after each cycle. MD simulations are performed in LAMMPS \cite{Plimpton:1995} while the MC moves, which are included to enhance the sampling of the preferred Au-S bonding geometries, are performed using an in-house code. The second-moment approximation to the tight-binding potential (TB-SMA) is used to describe Au-Au interactions; TB-SMA is a semiempirical, many-body potential that was developed by Cleri and Rosato \cite{Cleri:1993} to capture metallic bonding effects. Multi-site Morse bonding curves calibrated from density functional theory calculations are employed to describe S-Au bonding.\cite{Leng:2007} Van der Waals interactions between BDT and Au and intramolecular interactions within BDT are described using the universal force field.\cite{Rappe:1992} Electrostatic interactions are described using the Coulombic potential, with partial charges residing on BDT atoms derived from our prior work. \cite{Leng:2005} Further details are included in the Supporting Information.

\textit{Conductance Calculations}. Conductance calculations are performed using the geometries extracted directly from the hybrid MD/MC simulations; no geometry optimizations are performed within the DFT framework. The zero-bias conductance, given by $G = T(E_{F})G_{0}$, where $T$ is the transmission probability, $E_{F}$ is the Fermi level of the electrodes, and $G_{0} = 2e^{2}/h$, is calculated using SMEAGOL \cite{Rocha:2006,Rungger:2008}, an electron transport code that interfaces with the DFT package SIESTA \cite{Ordejon:1996,Sanchez-Portal:1997}. Self-interaction corrected DFT is employed to accurately describe the energy level lineup between the molecule and leads \cite{Toher:2008,Pontes:2011}. We obtain a conductance value of 0.06$G_{0}$ after optimizing the geometry of junction 1 (using the classical force fields), which matches the value reported in prior studies of ideal Au-BDT-Au junctions \cite{Toher:2008,Pontes:2011}, thus validating our force fields and the conductance calculations. This value of conductance (0.06$G_{0}$) along with most of the other values shown in Figure 1 are higher than the most-probable value (0.011$G_{0}$) reported in some experiments \cite{Xiao:2004,Tsutsui:2009}, while falling in the range of values measured in other experiments \cite{Kim:2011,Bruot:2012}. As pointed out in several previous studies \cite{DiVentra:2000,Muller:2006,Nichols:2010}, the long-standing discrepancy between the experimentally measured and theoretically calculated value may be due to shortcomings in the theory, structural differences between the experimental and theoretical junctions, or some other factor. Nevertheless, these differences are immaterial for the present work since the focus is on relative conductance differences rather than the exact values of conductance. Further details are included in the Supporting Information.

Note that we do not consider vibrational effects in the conductance calculations, instead focusing only on elastic transport processes. It has been shown \cite{Benesch:2006} that vibrational effects are negligible for BDT connected to Au tips, and for Au nanowires vibrational effects are small compared to the total transmission.\cite{Frederiksen:2007} We therefore expect a small vibrational contribution to the transport for our Au-BDT-Au structures.


\section{Acknowledgements}

WRF acknowledges partial support from the U.S. Department of Education for a Graduate Assistance in Areas of National Need (GAANN) Fellowship under grant number P200A090323; WRF, CRI and PTC acknowledge partial support from the National Science Foundation through grant CBET-1028374. IR, AMS, and SS thank the King Abdullah University of Science and Technology (ACRAB project) for financial support. This research used resources of the National Energy Research Scientific Computing Center (NERSC), which is supported by the Office of Science of the U.S. Department of Energy under Contract No. DE-AC02-05CH11231; specifically, the conductance calculations were performed on NERSC's Carver.

\noindent {\bf Supporting Information Available:} Transmission curves for junctions 1, 2, and 3; Details about the simulation methodology and conductance calculations. This material is available free of charge via the Internet at http://pubs.acs.org. 

%
%
%
\bibliography{nature-library}

\end{document}